\documentstyle[editedvolume,psfig,amssymb]{jd}
\def\Chris#1{{\bf[#1 -- Chris]}}

\def\Hyung-Mok#1{{\bf[#1 -- Hyung-Mok]}}
\def\hyung-mok#1{{\bf[#1 -- Hyung-Mok]}}

\newcommand{\msun}{\mbox{${\rm M}_\odot$}}

\newcommand{\ga}{\mbox{${\geq}$}}
\begin{opening}

\title{Stellar Evolution and Dynamics in Star Clusters}

\author{Simon F.\ Portegies Zwart}
\institute{Astronomical Institute {\em Anton Pannekoek}, 
                Kruislaan 403, NL-1098 SJ Amsterdam} 
\author{Christopher A. Tout}
\institute{Department of Mathematics, Monash University, Clayton,
Victoria 3168, Australia.}
\author{Hyung Mok Lee}
\institute{Dept. of Earth Sciences, Pusan National University, Pusan
609-735, Korea}
\end{opening}
\runningtitle{Stellar Evolution and Cluster Dynamics}
\begin{document}

\begin{abstract}
Dynamical models of star clusters are maturing in the sense that
effects other than simple point particle dynamics are taken into
account. We summarize the relevance of and prospects for this new
generation of $N$-body models.
\end{abstract}

\section{Summary and Introduction}
Realistic star clusters are not born instantly nor are they
isolated from external perturbations nor do
they consist of single time-independent
equal-mass points.
On the contrary a realistic star cluster is born gradually
from a contracting gas cloud, which is embedded in the external
potential of a galaxy, and consists of evolving single
stars as well as multiple systems.  Owing to lack of
computer power and partially to lack of software,
computations, in which all these effects are
accounted for, have not yet been performed.

However numerical models of star
clusters are beginning to come of age and incorporate deviations from
the ideal star cluster.  Some models already include mutual interaction
between stellar evolution and star cluster dynamics 
(see Chernoff \& Weinberg 1990; Fukushige \& Heggie 1995; 
de la Fuente Marcos 1996; Spurzem \& Aarseth 1996; Einsel \& Spurzem 1997;
Portegies Zwart et al.\ 1997; Tout et al.\ 1997, and the reviews Hut
et al.\ 1992 and Meylan \& Heggie 1997). 
Here we study the arguments for combining stellar
evolution and stellar dynamics in hybrid models, the advantages and
disadvantages of performing such model computations and
outline the future of dynamical models of star clusters.

\section{From zero to first order modelling}
Perturbations of the evolution of dynamical star clusters can be
subdivided into two classes: those which affect the cluster's
evolution from outside and are not significantly affected by the
evolution of the star cluster, and those which affect it from inside
and actually live in symbiosis with it undergoing mutual interaction.
This mutual interaction can be thought of as the ecology of the star
cluster (see Heggie 1992).

\subsection{The timescale argument}
The internal evolution of a stellar system is governed by two
fundamental time scales: these are the crossing time $t_{\rm crss} = r_{\rm
vir}/\langle v \rangle$, the ratio of the cluster virial radius to the
mean velocity of its components, and the relaxation time $t_{\rm
rlx}$.  The ratio of these $t_{\rm rlx}/t_{\rm crss} \equiv n_{\rm
rlx}$ is roughly proportional to $N$.  In real globular clusters $N
\approx 10^5$ and both timescales are usually well separated by more
than three orders of magnitude.

\begin{figure}
\centerline{
\psfig{file=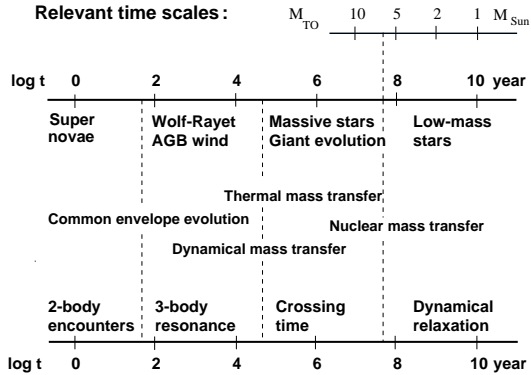,bbllx=550pt,bblly=60pt,bburx=45pt,bbury=770pt,height=5cm,angle=-90}}
\caption{
Time bar for the turn-off mass $M_{\rm TO}$ of star clusters
(upper bar), for the evolution of single stars (second bar) and for
the dynamical time scales of star clusters (lower bar).  Those for the
evolution of close binaries are indicated in between.  
}
\label{fig:timetable}
\end{figure}

Stimulated evolution of the star cluster such as stellar evolution
or the external potential of the Galaxy introduces new time scales.
It is generally the shortest timescale which drives the stimulated
evolution of the cluster and its stars so we must
model these complex systems with a sufficiently fine time
resolution.  Figure~\ref{fig:timetable} gives an overview of the
dominant timescales in a star cluster

The evolution timescale for low-mass stars ($M \lesssim 1\,\msun$)
exceeds the age of globular clusters so that the long term evolution
can be investigated with simple, few-component models containing
main-sequence stars, white dwarfs and neutron stars (Lee 1987, Kim et
al.\ 1998). The neutron stars, which are retained in these models,
strongly affect the dynamics of the stellar system; the mass of white
dwarfs is similar to that of normal stars while massive black holes
are easily ejected by the interaction with other cluster black holes
and binaries (Kulkarni et al.\ 1993).  These first-order models already
demonstrate the importance of the interaction between the dynamics of
the star clusters and its static stellar population. They are useful
for studying the future evolution of star clusters but their past
requires more refined models.

\section{Stimulated evolution}
Higher-order models incorporate the details of stimulated evolution
directly.
Processes which stimulate the evolution of the star cluster fall into
two categories, those which are affected by the internal
evolution of the host cluster and those which are not.

\subsection{External stimulation}
Star clusters are relatively low-mass entities in an orbit around the
much more massive Galaxy.  A globular cluster is embedded in
its effective Roche-lobe which limits the size of the cluster to its tidal
radius. The timescale for a galactic revolution is of the order of
$10^8$ to $10^9\,$yr.  Even if the galactocentric orbit is circular
the effect of the tidal field is, owing to the
characteristic pear shape of the Roche-lobe, not spherically symmetric.
The Coriolis force on the circulating cluster
and its fall through the galactic disc, twice per orbit,
are expected to affect the dynamical evolution of the star cluster
considerably.
These effects are relatively straightforward to
incorporate in dynamical models because there is no mutual
interaction (i.e.\ second order effects):
the cluster does not affect the evolution of the galactic disc.
Molecular clouds and neighbouring star clusters are possibly affected
by a passing cluster but this stimulated evolution is not coupled
back to the host cluster (Spurzem \& Giersz 1996).

\subsection{Internal stimulation}
A star cluster which consists of static equal-mass points
is only affected by its internal dynamical evolution
and external stimulated evolution. 
Once we relax the equal-mass assumption and use a more
realistic mass function we must also include stellar evolution.
In the past the mass function has been replaced by the equal-mass
approximation partly because of this requirement but also because
high-mass stars, which are considered to be the most interesting, are
very rare compared to the low-mass objects which absorb most of the
computer time.

The most massive stars ($m_\star \ga 10\,M_\odot$) evolve on a
timescale comparable to the crossing time of a globular cluster.  If
such stars maintain the same mass and radius throughout the dynamical
evolution of the stellar system, they sink to the cluster core by
dynamical friction (free-free scattering) on a timescale $t_{\rm df}
\propto t_{\rm rlx}{\langle m \rangle}/{m_\star}$ (of the order of $10^8$~yr)
and begin to dominate the dynamics there.  In reality such stars
lose most of their mass within a few crossing times.

\begin{figure}
\centerline{
\psfig{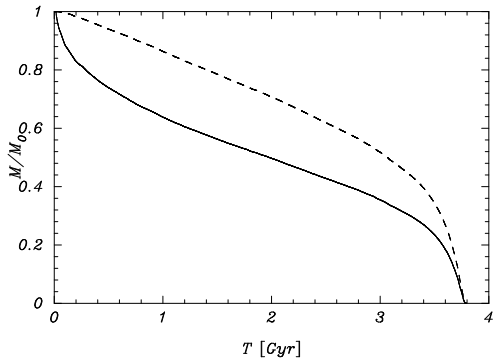}}
\caption{
The time evolution of the total mass (solid line) and number
of stars (dashed line) of the cluster for a
standard, 32768-body run. Both the mass and the number of stars are
normalized to their initial value (model M32 of Portegies Zwart et
al.~1998, {\em in prep}). 
The initial star cluster contains 32768 stars chosen from a power-law
initial mass function with exponent 2.5 (Salpeter = 2.35) between 0.4~\msun\
and 14~\msun. The cluster is situated in a circular orbit at a
distance of about 15~kpc from the Galactic centre.
}
\label{mass32krun}
\end{figure}

Figure~\ref{mass32krun} gives as a function of time the total mass and
the total number of stars (normalized to its initial value) for a
simulated star cluster.  The figure demonstrates the dominant effect of
mass loss by stellar evolution in the first few $10^8\,$yr
followed by a more gradual decrease of mass as the most massive stars
burn out and the mass loss is driven by stellar escapees. The final
epoch is dominated by escaping stars as the stellar system evaporates.

\section{Primordial binaries}
Primordial binaries are crucial for the dynamical evolution of a dense
star cluster; their binding energies are an important reservoir that
effectively heats the stellar system, preventing its core from collapsing.
Including them in the computations is crucial, but ignoring the
evolution of the binary components and together with that, the
evolution of the binary parameters, is unrealistic.  The binaries are
most strongly coupled to the evolution of the star cluster (Hut 1994)
and the variation in the binary parameters, owing to internal dynamical
evolution, can have a dramatic effect on the dynamical evolution of the
cluster as a whole.

The orbital parameters of binaries in clusters are affected by
internal evolution and by subsequent dynamical encounters; this may
lead to formation of objects which are difficult to explain by
evolution of isolated binaries.  The study of such stars
in globular clusters can provide observational signatures of the
relative importance of different dynamical and evolutionary
processes. We describe some of these signatures in the following
section.

\begin{figure}
\centerline{
\psfig{file=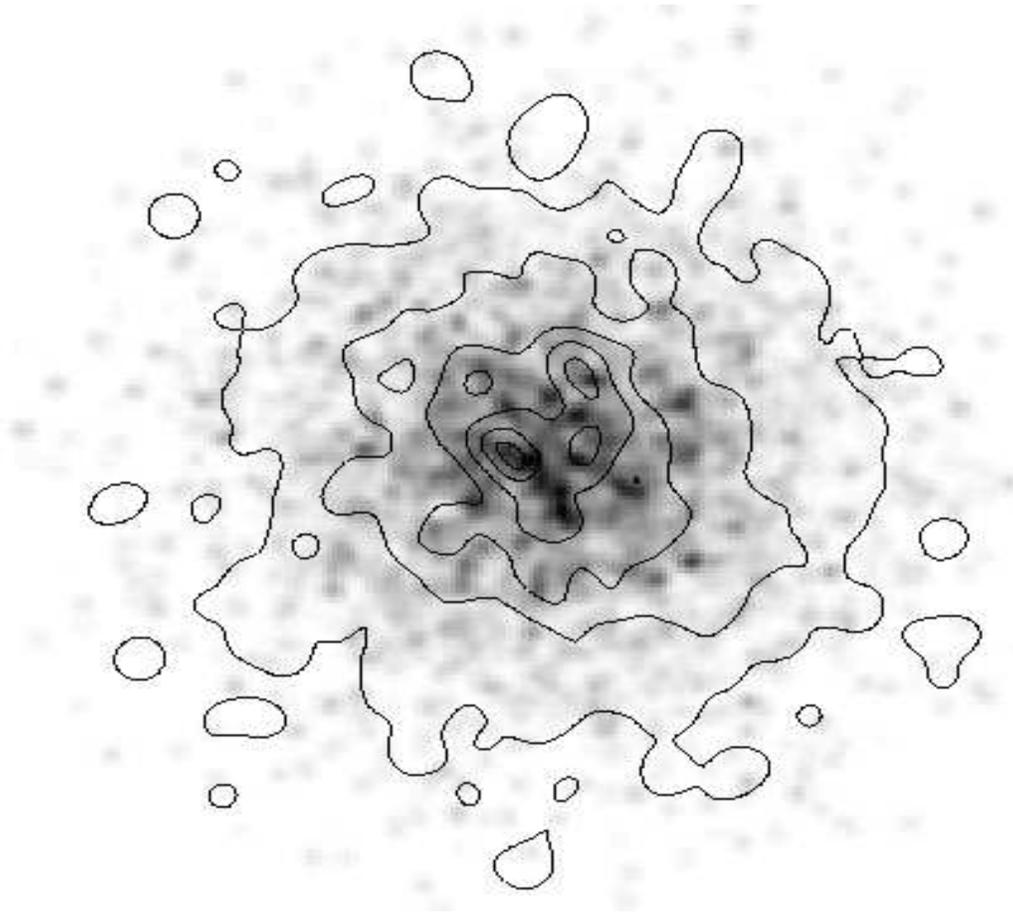,bbllx=50pt,bblly=75pt,bburx=425pt,bbury=550pt,height=12.5cm,width=12.5cm}}
\caption{ 
Picture of a synthesized globular stars cluster in visual
magnitude overlayed with a synthesized radio map.
Because the luminosity, temperature, angular size, mass etc.\ of all
stars in the simulation model are known it is straight forward to
produce a picture of the system as if observed by a telescope with
finite seeing.  The picture is taken from the star cluster in
fig.~\ref{mass32krun} at an age of $\sim 1\,$Gyr, at which point it
still contains about 28000 stars of which about 250 are neutron stars.
}
\label{ccd_M32}
\end{figure}

\section{Outlook for full models}

At present there are two $N$-body models which
incorporate such effects, Aarseth's {\tt NBODY5} (Tout et al.\ 1997)
and the hybridization of {\sf Kira} (Hut et al.\ 1995) and 
{\sf SeBa} (Portegies Zwart \& Verbunt 1996) in starlab.
Both these models have reached the end of their initial testing and
are ready for computation of realistic but small clusters
($<10^5\,$stars).  A direct
comparison of the two using the same initial conditions will be one of
the first major projects to be tackled.  Because these are full models
of both the position and velocity and the evolutionary state of each
individual stellar component it is possible to simulate observations
for comparison with real clusters.  Figure~\ref{ccd_M32} illustrates
how we can synthesize a ccd image of a distant cluster as seen with an
appropriate telescope.  With velocity information it is no more
difficult to generate a Doppler map.  In addition all the information
exists to follow the integrated luminosity and colours and
by including atmospheres for each star integrated cluster spectra can
be produced.  Because we follow the evolution of the whole cluster
over its entire lifetime we can
also calculate the evolution of any of these observable quantities
for any chosen spatial region.  Our only limitation at present is the
total number of stars we can include.  We are not yet able to compute
the dynamics of millions of stars in a realistic globular cluster
within a reasonable time.  Future generations of special purpose
computers (see Makino, these proceedings) will eventually make this
possible and we note here that there will be no problem including
detailed stellar evolution in such models because the time required
increases only linearly with the number of stars and interacting binaries.

These combined models allow the investigation of the detailed evolution of
individual stars and binaries, their mutual interaction with other
cluster members and the dynamical evolution of the stellar system as a
whole.  With the GRAPE series of computers (Ebisuzaki et al.~1994) it
is possible to tackle problems in which stellar evolution and stellar
dynamics are interactively connected.  We summarize a few specific
problems that can be addressed with these models:

\begin{itemize}
\item[$\bullet$]
{\em Birth rate of OB runaway stars.}

An OB runaway is a main-sequence star
which has an anomalously high space velocity compared to other O and B
stars in the galactic disc (Gies \& Bolton 1986). 
Two completely different scenarios have been proposed for their formation
in young clusters: 
the OB star is ejected from a close binary at the moment its companion
explodes as a supernova (Blaauw 1992); or
the OB runaway is ejected in a dynamical 3-body encounter (Leonard 1995).

\item[$\bullet$]
{\em The origin of X-ray binaries.}

A neutron star or a
black hole accretes from a Roche lobe filling, less massive companion.
The frequency of LMXBs in globular clusters
can be explained by tidal capture (Fabian et al.~1975),
but there are many doubts about the efficiency of this process (Ray 1987).
Similar mechanisms apply to
low-luminosity X-ray sources and cataclysmic variables,
which are preferentially found in the cores of massive
globular clusters with a high central density (Johnston \&
Verbunt 1996).
The discovery of close binaries in globular clusters has led
to an alternative formation mechanism, exchange
of a compact object into a pre-existing binary (Bailyn 1996).

\item[$\bullet$]
{\em Dynamical formation of triple systems.}

The millisecond pulsar PSR B1620-26 in the core of the globular
cluster M4 is possibly a triple system (Rasio et al.~1995). An
optical counterpart has been proposed by Bailyn et al.\ (1994); it is a
low-mass main-sequence star, which accompanies the neutron-star plus
white-dwarf binary.  Dynamical formation mechanisms are inevitable.

\item[$\bullet$]
{\em Colour gradients in globular-cluster cores.}

Post-collapse globular clusters tend to be bluer in the centre than in
the outskirts.  This colour gradient cannot be readily explained by
mass segregation alone.  An additional source of red stars in the
cluster halo together with an extra source of blue light in the
cluster centre are required to explain the observed colour gradients
(Djorgovski et al.~1991).  Related problems are the significant
depletion of subgiants in the cores of dense clusters and the
concentration of blue stragglers there when compared to horizontal
branch stars (Paresce et al.~1991).

\item[$\bullet$]
{\em Ejection of compact objects}

The concept of an intrinsic asymmetry in supernovae, ejecting the
newly formed compact object with a high velocity, is vividly debated
by upholders and antagonists from various fields of astronomy.  The
modest escape velocities of globular clusters make them excellent
laboratories for studying the kick velocity distribution.  The effect
of the presence or absence of compact objects affects the dynamical
evolution (Kim et al.\ 1997) and the binary population (Hut 1994) of a
star cluster considerably.

\item[$\bullet$]
{\em Formation of black holes in galactic nuclei.}

The apparent
central velocity dispersion in the cores of several dense star
clusters and the results from spectroscopic observations of nearby
galaxies suggests the presence of a massive central black hole
(Kormendy et al.~ 1996, Ford et al.~1994 and
Harms et al.~1994).  The formation of a central black hole
may result from runaway growth of a black hole following multiple
collisions with other cluster members
(Quinlan et al.~1995, Lee 1995). 

\end{itemize}

\acknowledgements This work was supported in part by the Netherlands
Organization for Scientific Research under grant PGS 13-4109, by
Spinoza grant 08-0 to E.~P.~J.~van den Heuvel, and by the Leids
Kerkhoven Boscha Fonds. SPZ thanks the Institute for Advanced Study
and the University of Tokyo for their hospitality.  
CAT is very grateful to Douglas Heggie and the UK PPARC for supporting
his attendance from
their HARP grant.

\end{document}